\documentclass[twocolumn,showpacs,preprintnumbers,amsmath,amssymb]{revtex4}
\usepackage{graphicx}% Include figure files
\usepackage{dcolumn}% Align table columns on decimal point
\usepackage{bm}% bold math

\def\thalf{{\textstyle{\frac{1}{2}}}}

\def\tquar{{\textstyle{\frac{1}{4}}}}

\def\ttquar{{\textstyle{\frac{3}{4}}}}

\def\s{\sigma}

\def\pv{\vmg{\pi}}

\def\Bmunu{\vmg{B_{\mu\nu}}}
\def\b{\vmg{b}}
\newcommand{\vm}[1]{\mbox{\bf#1}}
\newcommand{\vmg}[1]{\mbox{\boldmath$#1$}}

\newcommand{\bq}    {\begin{equation}}
\newcommand{\eq}    {\end{equation}}
\newcommand{\bqr} {\begin{eqnarray}}
\newcommand{\eqr} {\end{eqnarray}}
\begin{document}

\title{Softening of the equation of state of matter
at large densities and temperatures:\\ 
chiral symmetry restoration vs. quark deconfinement}

\author{Luca Bonanno}
\author{Alessandro Drago} 
\affiliation{
Dipartimento di Fisica, Universit\`a di Ferrara and 
INFN, Sez. Ferrara, 44100 Ferrara,
Italy} 
\author{Andrea Lavagno}
\affiliation{Dipartimento di Fisica, Politecnico di Torino and 
INFN, Sez. Torino, 10125 Torino, Italy}

\begin{abstract}

We discuss two models for describing the 
behavior of matter at large densities and intermediate temperatures.
In both models a softening of the equation of state
takes place due to the appearance of new degrees of freedom. The first 
is an hadronic model in which the softening is due to chiral symmetry
restoration.
In the second
model the softening is associated with the formation of 
clusters of quarks in the mixed phase. 
We show that both models allow a significant softening but,
in the first case the bulk modulus is mainly dependent on the density,
while in the mixed phase model it also strongly depends on the temperature.
We also show that the bulk modulus is not vanishing in the mixed phase due to
the presence of two conserved charges, the baryon and the isospin one.
Only in a small region of densities and temperatures the incompressibility
becomes extremely small. Finally we compare our results with
recent analysis
of heavy ion collisions at intermediate energies.
\end{abstract}


\maketitle

{\it Introduction.} --
The behavior of matter at large densities and temperatures
is still poorly known but, 
on general grounds,
new degrees of freedom are expected to appear and they should generate
a softening of the Equation of State (EOS).
In this Letter we 
discuss two models for the 
softening: the first model is based on a 
purely hadronic chirally symmetric EOS 
\cite{Heide:1993yz,Carter:1995zi,Carter:1997fn}
and the softening is
due to partial restoration of the chiral symmetry. In that model
the calculations are performed beyond mean field approximation
and quantum fluctuations play a crucial role.
In a second model the
softening is due to the 
formation of clusters of quarks, which are the precursors
of deconfinement. These clusters are at first metastable and
they stabilize only at larger densities.

The aim of our calculation is to compare how the softening takes 
place in the two models and finally to relate our results to recent analysis 
of the experimental data. This may be helpful also in prevision of future 
experiments planned e.g. at facility FAIR at GSI \cite{Senger:2004jw}.

The extraction of experimental information about the EOS
of matter at large densities and temperatures from the data of 
intermediate
and high energy Heavy Ion Collisions (HICs) is very complicated. 
Possible indirect indications of a softening of the EOS at the energies 
reached at AGS have been discussed several times in the literature 
\cite{Liu:1998yc,Soff:1999yg,Sahu:2002ku,Stoecker:2004qu,Stoecker:2004xc,Isse:2005nk}.
In particular, a recent analysis \cite{Russkikh:2006ae} based on a 3-fluid
dynamics simulation suggests a progressive softening of the
EOS tested through HICs at energies ranging from 2$A$ GeV up to
8$A$ GeV.

In this letter we will show that in the chiral
model the incompressibility strongly depends on the density but it is almost
temperature independent at least up to $T\sim$ 150 MeV. At the contrary a 
strong temperature dependence is found in the mixed phase model.
Since in HICs at intermediate energies not too large densities
are reached, the mixed-phase model seems to provide a better description 
of the results of the experiments analysis.

{\it The Chiral-dilaton model.} --
It is not trivial to develop chiral invariant models in which
saturation properties of nuclei are well reproduced.
It is well known that the attempt of using e.g. the sigma model
to describe nuclear dynamics fails due to the impossibility to
reproduce basic properties of nuclei \cite{Furnstahl:1995zb}.
More sophisticated approaches have been proposed in the literature,
both within a SU(2) chiral symmetric model 
\cite{Mishustin:1993ub,Furnstahl:1995by} and also 
extending the symmetry to the strange sector 
\cite{Papazoglou:1997uw,Papazoglou:1998vr,Wang:2003cn}.
Here we use the model introduced by the Minnesota group
\cite{Heide:1993yz,Carter:1995zi,Carter:1997fn}.
In that model chiral fields are present together with 
a dilaton field 
which reproduces at a mean
field level the breaking of scale symmetry which takes place in QCD.
In \cite{Heide:1993yz,Carter:1995zi,Carter:1997fn} 
it has been developed a formalism (which we adopt) 
allowing resummations beyond mean field approximation. This is
important when studying a strongly non-perturbative problem as
the restoration of chiral symmetry. The lagrangian of the model reads:

\begin{eqnarray}
{\cal L}&=&\thalf\partial_{\mu}\sigma\partial^{\mu}
\sigma+\thalf\partial_{\mu}\vmg{\pi}\cdot\partial^{\mu}\vmg{\pi}
+\thalf\partial_{\mu}\phi\partial^{\mu}\phi
-\tquar\omega_{\mu\nu}\omega^{\mu\nu}\nonumber\\
&-&\tquar\Bmunu\cdot\Bmunu
+\thalf G_{\omega\phi}\phi^2 \omega_\mu\omega^\mu 
+\thalf G_{b\phi}\phi^2 \b_\mu\cdot\b^\mu\nonumber\\ 
&+&[(G_4)^2\omega_\mu\omega^\mu]^2-{\cal V}\\
&+&\bar{N}\left[\gamma^\mu(i\partial_{\mu}-g_\omega\omega_\mu
-\thalf g_{\rho}\b_{\mu}\cdot\vmg{\tau})
-g\sqrt{\s^2+\pv^2} \right] N \label{lb}\nonumber
\end{eqnarray}
where
\begin{eqnarray}
{\cal V}&=& B\phi^4
\left(\ln\frac{\phi}{\phi_0}-\frac{1}{4}\right)
\hspace{-.73mm}-\hspace{-.73mm}\thalf B\delta\phi^4
\ln\frac{\sigma^2+\vmg{\pi}^2}{\sigma_0^2}\nonumber\\
&+&\hspace{-.74mm}\thalf B\delta \zeta^2\phi^2\!\!\left[\sigma^2
+\vmg{\pi}^2-\frac{\phi^2}{2\zeta^2}\right]-\ttquar\epsilon_1'\label{lm}\\
&-&\tquar\epsilon_1'\left(\frac{\phi}{\phi_0}\right)^{\!2}
\left[\frac{4\sigma}{\sigma_0}-2\left(\frac{\sigma^2
+\vmg{\pi}^2}{\sigma_0^2}\right)-\left(\frac{\phi}{\phi_0}\right)^{\!2}
\,\right]\,.  \nonumber
\end{eqnarray}
Here $\sigma$ and $\vmg{\pi}$ are the chiral fields, $\phi$ the
dilaton field, $\omega_{\mu}$ the vector meson field and ${\bf
b}_{\mu}$ the vector-isovector meson field, introduced in order to study 
asymmetric nuclear matter. The field strength
tensors are defined in the usual way
$F_{\mu\nu}=\partial_{\mu}\omega_{\nu}-\partial_{\nu}\omega_{\mu}$,
$\vm{B}_{\mu\nu}=\partial_{\mu}\vm{b}_{\nu}-\partial_{\nu}\vm{b}_{\mu}$.
In the vacuum ${\phi}={\phi_0}$, ${\sigma}={\sigma_0}$ and
${\vmg{\pi}}=0$.  The $\omega$ and $\rho$ vacuum masses are generated
by their couplings with the dilaton field so that
$m_{\omega}=G_{\omega\phi}^{1/2} \phi_0$ and
$m_{\rho}=G_{\rho\phi}^{1/2} \phi_0$. Moreover
$\zeta={\phi_0}/{\sigma_0}$, B and $\delta$ are constants and
$\epsilon_1'$ is a term that breaks explicitly the chiral invariance
of the lagrangian.  The potential ${\cal V}$ of Eq.(\ref{lm}) 
is responsible for the scale symmetry breaking. The
choice of such a potential comes from the necessity to reproduce the
same divergence of the scale current as in QCD.
In our calculation we use the parameters set of Ref.\cite{Carter:1997fn}
which was able to reproduce nuclear spectroscopy and
also gives, within this model, the smallest value for the incompressibility at 
saturation density, $K^{-1}$ = 322 MeV. Notice that this
value is slightly larger than those traditionally used.

\begin{figure}
\includegraphics[scale=0.4]{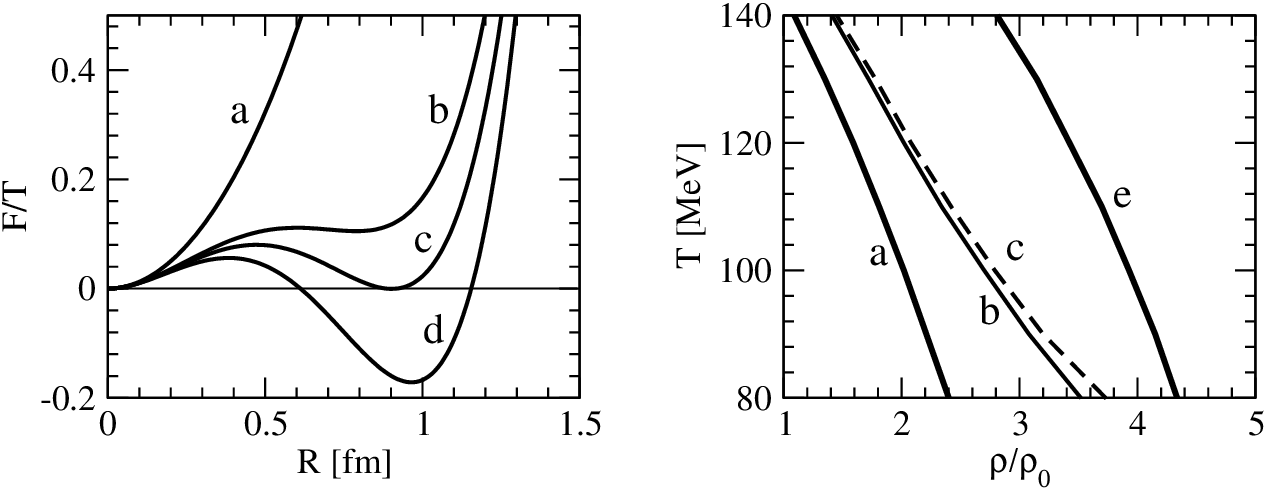}
\caption{\label{mixed_phase}
The left panel represents the free energy of a bubble of quarks as a
function of its radius R, while in the right panel the temperature
vs. the baryon density is shown. The $a$ and $e$ curves correspond to the
critical densities delimiting the mixed phase 
in the absence of surface tension; the densities
between $b$ and $c$ correspond to the formation of a metastable
bubble. For densities larger than $c$ the mixed phase is stable 
(an example of the corresponding free energy is provided by the $d$ curve).}
\end{figure}

{\it The mixed hadron-quark phase.} --
If the deconfinement transition at finite density is first order,
a mixed phase can form and it is typically described using two separate
EOSs, one for the hadronic and one for the quark phase.
Concerning the hadronic phase we use a relativistic field theoretical model,
the NL$\rho\delta$ \cite{Liu:2001iz}, 
taking into account also a scalar-isovector interaction, which increases
the symmetry energy only at large densities. Qualitatively similar
results can be obtained using other hadronic models.  For the quark
phase we adopt an MIT bag like model. It is well known that, using the
simplest version of the MIT bag model, if the bag pressure $B$ is
fixed to reproduce the critical temperature computed in lattice QCD,
then, at moderate temperatures, the deconfinement transition takes place
at very large densities.  On the other hand there are strong
theoretical indications that at moderate and large densities (and not
too large temperatures) diquark condensates can form, whose effect can
be approximately taken into account by reducing the value of the bag
constant.  A phenomenological approach can therefore be based on a
density dependent bag constant, as proposed in
Refs.\cite{Burgio:2002sn,Grigorian:2003vi}. We have adopted a
parametrization of the form 
\bq B_{\mathrm{eff}}=B_0-[\Delta(\mu)]^2 \mu^2 \,, \eq 
where $\Delta(\mu)=\bar{\Delta}
\,{\mathrm{exp}}[-(\mu-\mu_0)^2/a^2]$.  Here ${B_0}^{1/4}=215$
MeV, $\bar{\Delta}=100$ MeV, $\mu_0=300$ MeV and $a=300$ MeV. 
One gluon exchange corrections are taken into account and 
we use $\alpha_s=0.35$. 
One constraint on the parameter values is that at $\mu=0$ the critical
temperature is $\sim$ 170 MeV, as suggested by lattice calculations,
while the other constraint is the requirement that the mixed phase
starts forming at a density slightly exceeding 3$\,\rho_0$ for a
temperature of the order of 90 MeV (as also suggested e.g. 
by \cite{Toneev:2003sm}). The choice of the
parametrization is clearly inspired by the results of microscopical
analysis on diquark condensate (for recent reviews see
e.g. \cite{Ruster:2006aj,Alford:2006fw}). On the other hand, the
actual formation of a condensate and, even more, its specific type are
still very uncertain
\footnote{The possible condensate would be build on u and d
quarks. It is uncertain if such a condensate can exist at
temperatures of several ten MeV, if it can form in partially isospin
asymmetric matter and if it produces a stable state.}, what
justifies a phenomenological approach as ours.

To describe the mixed phase we use the Gibbs formalism, which in
Refs.~\cite{Glendenning:1992vb,Muller:1995ji,Muller:1997tm} has been applied to
systems where more than one conserved charge is present. In this
Letter we are studying the formation of a mixed phase in which both
baryon and isospin charge are preserved 
\footnote{In Ref.\cite{Toneev:2003sm} the same formalism has
been adopted, but the conserved charges were the baryonic 
one and strangeness.}.
The main result of this
formalism is that, at variance with the so called Maxwell construction,
the pressure is {\it not} constant in the mixed phase and therefore
the incompressibility does {\it not} vanish \footnote{From the viewpoint of
Ehrenfest's definition, the transition with two conserved charges
is not of first, but of second order \cite{Muller:1995ji,Muller:1997tm}.}.
%qui c'era la figura 1

An important issue concerns the effect of a surface tension at the
interface between hadrons and quarks. The value of such a tension is
poorly known, and in an MIT-bag-like model it is dominated by the
effect of finite masses \cite{Voskresensky:2002hu}, in particular of
the strange quark. On the other hand in the system studied here
strangeness plays a minor role, if any, and therefore the surface
tension $\sigma$ should be rather small. In the analysis we have used
$\sigma$ = 10 MeV/fm$^2$, but the results are qualitatively similar if
a slightly larger value of $\sigma$ is used. In the left panel of
Fig.~\ref{mixed_phase} we show the free energy for the nucleation of a
bubble of quarks for various values of the density, computed using the
formalism of Ref.~\cite{DiToro:2006pq}. One can notice that the
density region in which quark bubble nucleation can really form
shrinks with respect to the case with $\sigma=0$. Moreover, there is an
intermediate density region in which bubbles can form, but they are
metastable. Only at larger densities a stable mixed phase can form. In
our calculation this region is rather tiny, as shown in the right
panel of Fig.~\ref{mixed_phase}, but its size would increase for
larger values of $\sigma$. This region of metastable quark bubbles is
conceptually very interesting, because it provides a link with the
so-called hadron-string-dynamics model \cite{Cassing:2000bj}. In both
descriptions, metastable matter is produced in an intermediate density window.
In the string formation scenario that matter
is made of unstable hadrons while in the scenario here presented it is made
of (small) unstable bubbles of quarks.

\begin{figure}[t]
\includegraphics[scale=0.3]{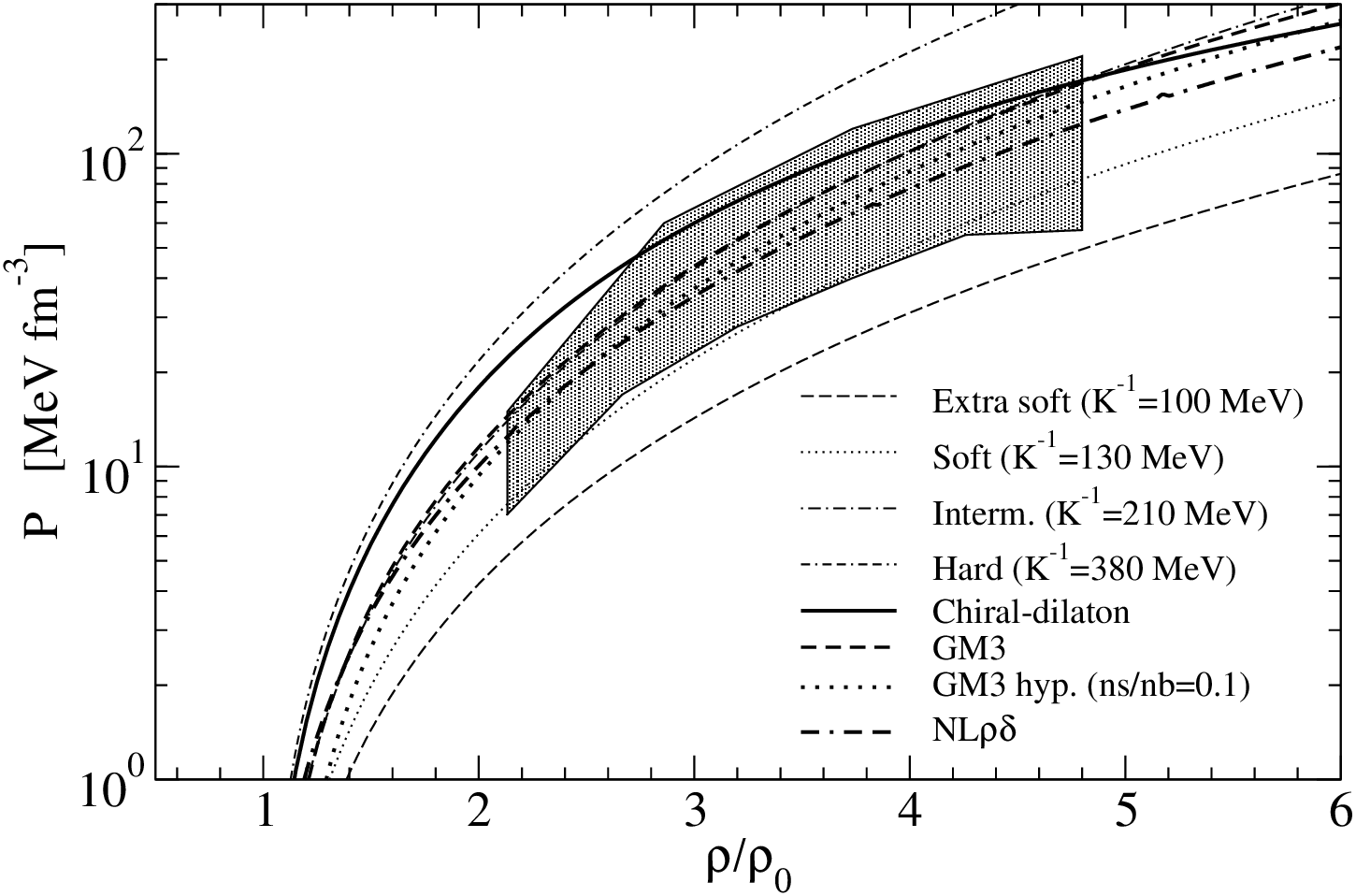}
\caption{\label{pressione}
Pressure at T=0 for symmetric nuclear matter, 
obtained using the chiral-dilaton model, the NL$\rho\delta$
model, the GM3 model and the set of simple parametric EOSs
used in Ref.~\cite{Russkikh:2006ae}. Also the pressure obtained using
the GM3 with a small hyperon fraction is shown.}
\end{figure}

{\it Results.}--
We start by comparing in Fig.~\ref{pressione} the pressure computed
using the models here discussed with the limits obtained from the
analysis of HICs at intermediate energies \cite{Danielewicz:2002pu}.
We have also introduced an EOS with a fixed fraction of strangeness due to
the presence of hyperons (using the GM3 parametrization).
In HICs a small number of hyperons is generated through 
associated production. An estimate of the number of hyperons
per participant can be obtained from the experimental ratio
of the kaon yields per participant \cite{Ahle:1999va},
which indicates a strangeness fraction smaller than $10\%$
\footnote{Therefore in the analysis of the mixed phase we have neglected
the strangeness content.}.
As a reference, we also show the parametric EOSs
discussed in \cite{Russkikh:2006ae}, which are characterized by different 
values of the 
incompressibility parameter $K^{-1}$.
From Fig.~\ref{pressione}
it is clear that only the extra soft and the hard EOSs are excluded
by the experimental limit. The pressure computed using the
chiral-dilaton model marginally exceeds the limit at low densities,
due to the too large value of the incompressibility at saturation density. 
The effect of the partial restoration of 
chiral symmetry is clearly visible as a softening taking place
at larger densities. Instead the small fraction of strangeness 
introduced using the GM3 parametrization does not produce a  
sizable softening of the EOS.

\begin{figure}[!t]
\includegraphics*[scale=0.42]{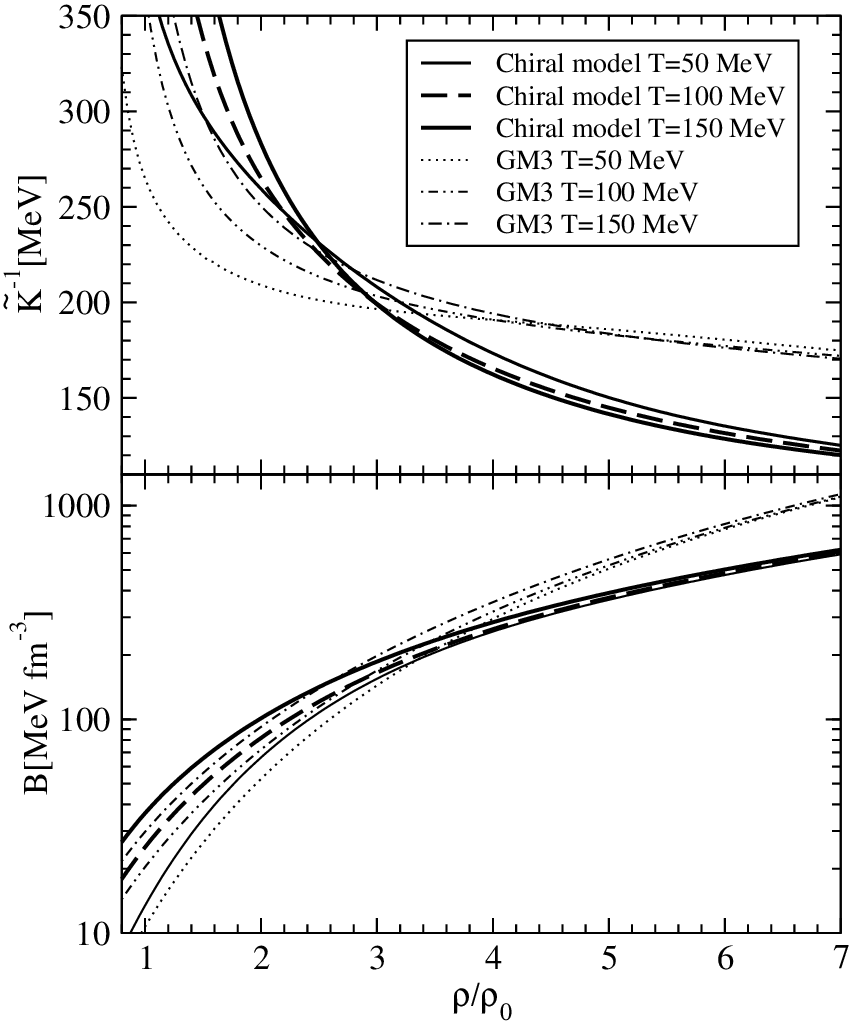}
\caption{\label{compressibilita}
The upper and lower panels show respectively the parameter
$\widetilde{K}^{-1}$ and the bulk modulus at different temperatures 
as a function of density for
the chiral-dilaton model and the GM3 parametrization. Here we used $Z/A=0.4$.}
\end{figure}

\begin{figure}[!t]
\includegraphics*[scale=0.42]{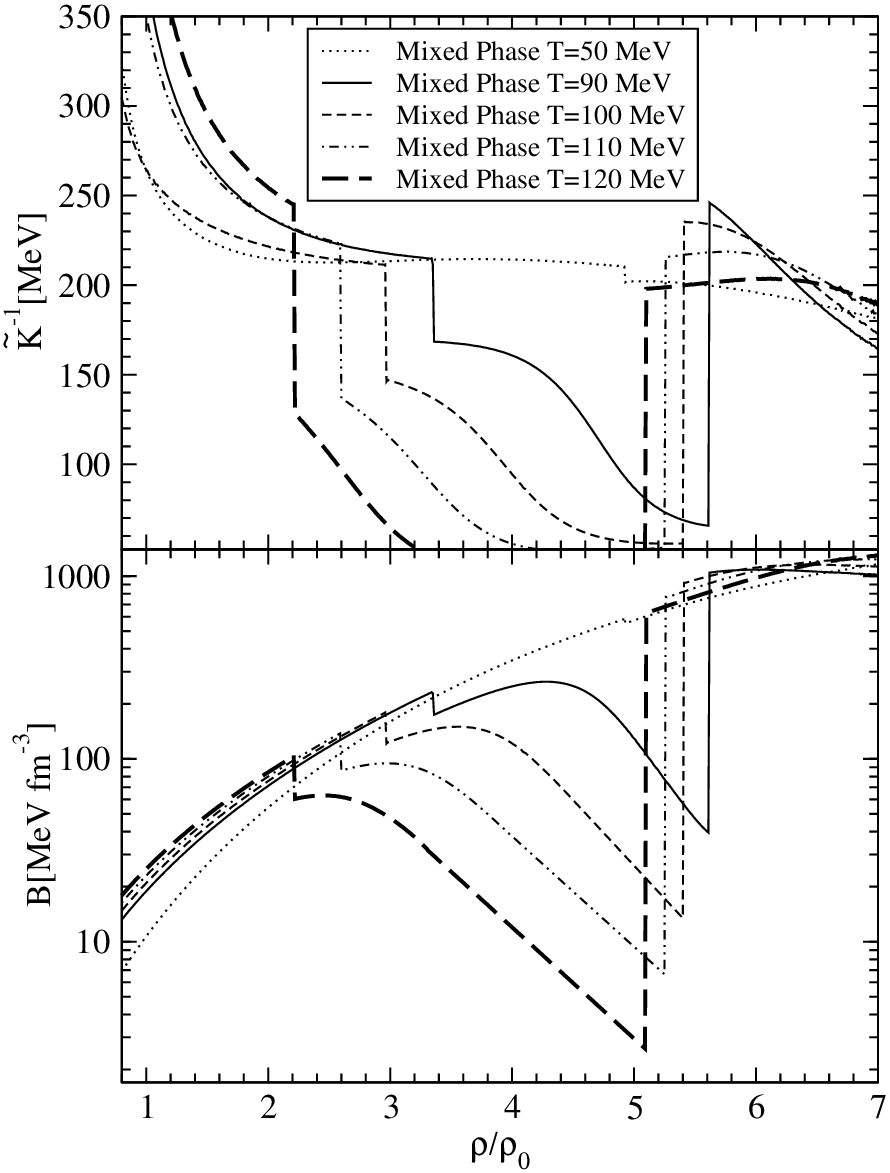}
\caption{\label{compressibilita2}
Same as in Fig. 3, but using the mixed-phase model.}
\end{figure}

We now compare the compressibility computed in the various models
discussed above with the one estimated in \cite{Russkikh:2006ae}. 
In the following we always assume that $Z/A=0.4$. In
the lower panels of Figs.~\ref{compressibilita} 
and~\ref{compressibilita2} we plot the bulk
modulus $B=\rho\,\partial P/\partial\rho$, 
respectively for the chiral dilaton model and for the mixed-phase model. 
Similarly, in the upper panels we show, as a function of the
density and temperature, 
the value of the incompressibility parameter for which
the bulk modulus of a parametric EOS has the same value of the
bulk modulus computed in our models. This quantity, called 
$\widetilde{K}^{-1}$
is obtained by solving the equation:
\begin{equation}
\left.\frac{\partial P_{par}(\tilde K)}{\partial\rho}\right\vert_{\rho,T}=
\left.\frac{\partial P_{model}}{\partial\rho}\right\vert_{\rho,T}.
\end{equation}
In this way we can directly compare with
the analysis of Ref.\cite{Russkikh:2006ae} which explicitly indicates
various values of the parameter $\widetilde{K}^{-1}$ as representative
of the EOS tested at various energies. In the
upper panel of Fig.~\ref{compressibilita} we show
the $\widetilde{K}^{-1}$ parameter computed using the
chiral-dilaton model and we see that it decreases significantly at large
densities, while its dependence on the temperature is relevant only at
small densities. By comparison, the GM3 parametrization provides a  
$\widetilde{K}^{-1}$ which is roughly constant at large densities.

In Fig.~\ref{compressibilita2}, we show that
the incompressibility computed using the mixed-phase model 
remains rather large above the lower critical
density and it becomes really small only approaching the upper
critical density.  It is extremely interesting to notice the strong
dependence of the incompressibility on the temperature. For instance,
at T=50 MeV and $\rho=3 \rho_0$ we still obtain
$\widetilde{K}^{-1}\sim 220$ MeV, but at T=90 MeV and $\rho=3.4
\rho_0$ we obtain $\widetilde{K}^{-1}\sim 170$ MeV and at T=100 MeV
and $\rho=3.6 \rho_0$ then $\widetilde{K}^{-1}\sim 120$ MeV.  The
density-temperature region where the incompressibility is extremely small
is actually rather limited, because for $T\gtrsim$ 150 MeV and
$\rho\gtrsim 4.5 \rho_0$ the pure quark matter phase is reached, for
which the incompressibility is large again. Notice also that, while at
lower temperatures the values of the incompressibility is rather
large, when T exceeds $\sim 120$ MeV the incompressibility is so small
that, for practical purposes, the pressure can be assumed to be
constant.

A possible scenario based on the mixed-phase formation is the following.
At 40A GeV the collapse of the elliptic flow has been interpreted as
a signal of a first order phase transition to quark-gluon plasma
\cite{Stoecker:2004qu,Stoecker:2004xc}.
At lower energies, a mixed phase is produced, but due to the existence of two
conserved charges 
%the transition is of second order. 
the pressure is not constant
and therefore dramatic manifestations of the formation of a mixed-phase are not 
expected. For instance, no wiggle in the behavior of the directed flow
vs. rapidity is expected,
since the appearance of wiggles is related to a strong reduction of
the incompressibility 
\cite{Csernai:1999nf}. Only rather sophisticated analysis,
as e.g. the extraction of the bulk modulus, can reveal the formation
of a mixed phase.
It will be interesting to test in future experiments
if the oscillatory behavior
of the directed flux, observed at 40A GeV, does
gradually take place at energies exceeding $\sim 10$A GeV
\cite{Senger:2004jw}, as suggested by our analysis.

In conclusion, a significant softening of the EOS can be
obtained either via chiral symmetry restoration, if large densities
are reached, or via the formation of a mixed phase of quarks and hadrons. 
The gradual reduction of the bulk modulus discussed in our analysis applies
to all situations in which the mixed phase forms at not too large
energy densities, as e.g. in the scenarios discussed in
Refs.~\cite{Arsene:2006vf,Klahn:2006iw}.

It is a pleasure to thank
D. Blaschke, P. Danielewicz, Y.B. Ivanov, B.A. Li,
J. Schaffner-Bielich and P. Senger
for useful discussions.

\bibliography{biblio}
\bibliographystyle{apsrev}
%\bibstyle{h-physrev3}

\end{document}